\def\QRCArxivVersion{1}
\documentclass[10pt,conference]{IEEEtran}
\IEEEoverridecommandlockouts
\newif\ifQRCauthorversion
\newif\ifQRCarxivversion
\QRCauthorversiontrue
\ifdefined\QRCAnonymousVersion
  \QRCauthorversionfalse
\fi
\ifdefined\QRCArxivVersion
  \QRCarxivversiontrue
\fi
\usepackage{cite}
\usepackage{amsmath,amssymb}
\usepackage{graphicx}
\usepackage{booktabs}
\usepackage{xcolor}
\usepackage{microtype}
\definecolor{repoLinkBlue}{RGB}{0,92,184}
\ifQRCarxivversion
  \usepackage[hypertexnames=false]{hyperref}
  \hypersetup{
    colorlinks,
    linkcolor={red!80!black},
    citecolor={green!80!black},
    urlcolor={blue!80!black}
  }
  \newcommand{\QRCdoi}[2]{\href{#1}{#2}}
  \newcommand{\QRCemail}[1]{\href{mailto:#1}{#1}}
  \newcommand{\QRCprojectlink}{\href{https://github.com/eybmits/qrc-operating-band}{\textcolor{repoLinkBlue}{project repository}}}
\else
  \newcommand{\QRCdoi}[2]{#2}
  \newcommand{\QRCemail}[1]{#1}
  \newcommand{\QRCprojectlink}{repository at github.com/eybmits/qrc-operating-band}
\fi
\newcommand{\orcid}[1]{}
\ifQRCauthorversion
  \ifQRCarxivversion
    \usepackage{orcidlink}
    \renewcommand{\orcid}[1]{\,\orcidlink{#1}}
  \fi
\fi
\ifQRCarxivversion
  \makeatletter
  \let\QRC@original@IEEEtitlepagestyle\ps@IEEEtitlepagestyle
  \def\ps@IEEEtitlepagestyle{%
    \QRC@original@IEEEtitlepagestyle
    \def\@oddfoot{\hbox to\textwidth{%
      \parbox[b]{\textwidth}{\raggedright\fontsize{5}{5.8}\selectfont
        \textcopyright{} 2026 IEEE. Personal use of this material is permitted. Permission from IEEE must be obtained for all other uses, in any current or future media,\\
        including reprinting/republishing this material for advertising or promotional purposes, creating new collective works, for resale or redistribution to servers or lists,\\
        or reuse of any copyrighted component of this work in other works.}\hss}}%
    \let\@evenfoot\@oddfoot
  }
  \makeatother
\fi
\newcommand{\PhaseBTwentyQSeventySize}{22}

\newcommand{\PhaseBTwentyQSixtySize}{41}
\newcommand{\PhaseBThirtyQSeventySize}{70}

\newcommand{\PhaseRepresentativeGamma}{0.22}
\newcommand{\PhaseLotoMeanRank}{0.193}
\newcommand{\PhaseLotoCILow}{0.176}
\newcommand{\PhaseLotoCIHigh}{0.210}
\newcommand{\PhaseLosoMeanRank}{0.151}
\newcommand{\PhaseLosoCILow}{0.139}
\newcommand{\PhaseLosoCIHigh}{0.163}

\newcommand{\PhaseHoldoutRows}{Mackey--Glass & 7.05e-05 & 5.26e-05 & 0.104\\
Lorenz & 1.51e-05 & 1.52e-05 & 0.136\\
NARMA & 0.560 & 0.607 & 0.204\\
Sunspots & 0.209 & 0.209 & 0.168\\
\textbf{All} & 0.192 & 0.204 & 0.153\\}

\newcommand{\PhaseMCSpearman}{-0.88}

\begin{document}
\raggedbottom
\title{Where a Quantum Reservoir Works: A Transferable Operating Band}
\ifQRCauthorversion
\author{%
\IEEEauthorblockN{%
Markus Baumann\IEEEauthorrefmark{1}\orcid{0009-0007-3575-1006}\thanks{Corresponding author: \QRCemail{markus.baumann@campus.lmu.de}.},
Itamar Fink\IEEEauthorrefmark{1}\orcid{0009-0000-2262-4445},
Johannes Wittmann\IEEEauthorrefmark{1}\orcid{0009-0009-9240-7257},
Claudia Linnhoff-Popien\IEEEauthorrefmark{1}\orcid{0000-0001-6284-9286},
and Jonas Stein\IEEEauthorrefmark{1}\orcid{0000-0001-5727-9151}}
\IEEEauthorblockA{\IEEEauthorrefmark{1}\textit{QAR-Lab, Department of Computer Science, LMU Munich, Munich, Germany}}
}
\else
\author{\IEEEauthorblockN{Anonymous Authors}}
\fi
\maketitle

\begin{abstract}
In quantum reservoir computing, a fixed quantum system transforms an input signal, while learning reduces to training a simple linear readout on its measured outputs. Since the quantum dynamics themselves are never optimized, the method is well suited to today's hardware. Yet these dynamics must still be chosen carefully, because their settings remain fixed throughout training and inference. It therefore remains open whether useful dynamics occupy a task-transferable region of control space and whether that region can be found without target-task tuning. We address this question for a dissipative reservoir by mapping performance over three central physical controls: the strength of the input drive, the coupling between neighboring qubits, and the rate of dissipation. Good performance concentrates in a well-defined operating band of this control space. This region transfers across tasks and reservoir initializations, and the same regime persists under an architectural change. It is also mechanistically grounded, since it disappears whenever any of the mechanisms that create it is removed. Finally, the region can be located cheaply before any task is run, using a simple memory diagnostic.
\end{abstract}

\begin{IEEEkeywords}
quantum machine learning, reservoir computing, quantum computing, dissipative systems, memory capacity
\end{IEEEkeywords}

\begin{figure*}[t!]
\centering
\includegraphics[width=\textwidth]{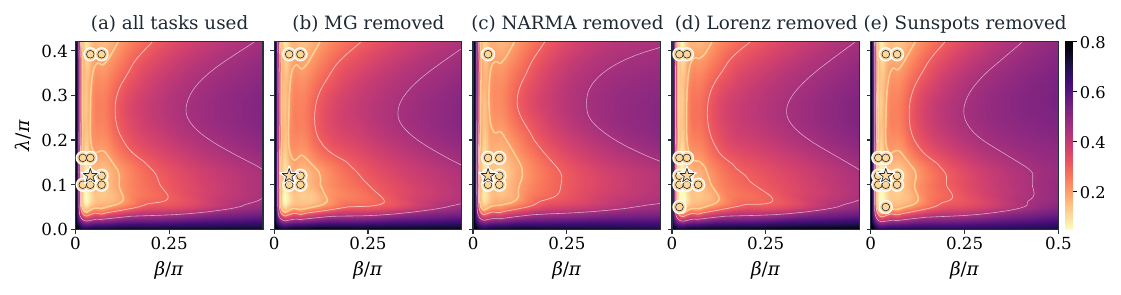}
\caption{\footnotesize The good operating region is essentially the same whichever task we score. Each panel maps validation rank over input drive $\beta$ and coupling $\lambda$ at fixed damping $\gamma = 0.12$. Brighter means better rank. Panel~(a) uses all four tasks; panels~(b) to (e) each drop one. Gold circles mark settings reaching top-quintile rank in at least $70\%$ of the displayed task--seed pairs; the star marks the full band's medoid projected onto $(\beta,\lambda)$. The basin and circles barely move when a task is dropped, so the band is not an artifact of any single task.}
\label{fig:phase}
\end{figure*}
\section{Introduction}
Reservoir computing builds on a simple principle: a fixed dynamical system, left untrained, transforms an input signal into a high-dimensional record of its recent history, and learning reduces to fitting a single linear readout on top of that record \cite{maass2002,jaeger2004,tanaka2019}. This simplicity makes the approach appealing for near-term quantum hardware: native quantum evolution can serve as the reservoir, and only the classical readout must be trained \cite{fujii2017,chen2020}. At the same time, it places the full burden on the fixed dynamics, raising a concrete design question: is there a task-transferable region of useful dynamics, and can it be found without target-task tuning?

Recent QRC work spans noisy gate-model, photonic, and continuous-variable platforms, and protocols from weak measurement to engineered dissipation and optical memory control \cite{chen2020,nokkala2021,mujal2023,sannia2024,paparelle2026}. Across this range the question persists, and classical reservoir theory already supplies the warning: maximal expressivity alone does not make a reservoir useful. A reservoir works by holding a balance, retaining recent input, forgetting old input in a controlled way, and exposing nonlinear functions of that history to the readout \cite{dambre2012,tanaka2019}. A quantum reservoir inherits this lesson but complicates it, because these ingredients are no longer free dials. They are consequences of distinct physical resources: how strongly each input perturbs the carried state, how coherently the system mixes that state across its parts, and how quickly it is allowed to forget \cite{chen2020,sannia2024}. Closest to our question, dynamical-phase analyses of Hamiltonian spin-network reservoirs locate good operation in the thermal phase near the thermalization transition \cite{martinez2021}. We address the complementary dissipative, gate-model setting by mapping its physical controls, testing transfer on withheld tasks, seeds, and time, ablating each mechanism, and evaluating task-free screening.

For a stateful dissipative gate-model reservoir, we treat this three-resource space as the proper setting for asking where the reservoir works, and we map it. What emerges is not a single tuned operating point but a stable operating band, characterized by gentle input, nonmaximal coupling, and controlled damping, whose boundaries stay essentially unchanged from task to task. The contribution of this paper is that region itself: (i) its existence as a stable operating band; (ii) its transfer across tasks, reservoir initializations, and a chronological holdout; (iii) its mechanistic grounding, since every single-ingredient ablation destroys it; and (iv) its predictability, since a task-free memory diagnostic locates it before any task-specific search.

\section{Protocol}

\subsection{Reservoir}
The reservoir is a small qubit system run as a recurrent loop. At each step it receives one scalar input value, evolves, and hands its state forward without ever being reset, so the carried state holds a record of recent inputs. We write $\rho_{t,\ell}$ for the state after input $t$ has passed through layer $\ell$; with $L$ the final layer, $\rho_{t-1,L}$ is handed forward from the previous step.

At step $t$, the input $u_t$ is written into this carried state by the same $Y$ rotation on every qubit,
\begin{equation}
\rho_{t,0}=U_{\rm in}(u_t)\,\rho_{t-1,L}\,U_{\rm in}^\dagger(u_t),
\qquad U_{\rm in}(u_t)=\bigotimes_i R_y(\beta u_t),
\end{equation}
with $\beta$ the input drive.

The perturbed state then passes through $L$ identical layers, each a frozen local mixer $U_{\rm mix}$, a ring $ZZ$ coupling $U_\lambda$, and amplitude damping $D_\gamma$ applied to every qubit,
\begin{equation}
\begin{aligned}
\rho_{t,\ell}
&=D_\gamma^{\otimes n}\!\left[U_\lambda U_{\rm mix}\,\rho_{t,\ell-1}\,U_{\rm mix}^\dagger U_\lambda^\dagger\right],\\
U_\lambda
&=\prod_{(i,j)\in E_{\rm ring}} e^{-i\lambda Z_iZ_j},
\end{aligned}
\end{equation}
with $E_{\rm ring}$ the nearest-neighbor edges of the qubit ring, $\lambda$ the coupling strength, and $\gamma$ the damping rate. Everything else is frozen: we vary only $\theta=(\beta,\lambda,\gamma)$, with $\beta$ and $\lambda$ angles in radians and $\gamma\in[0,1]$ the amplitude-damping probability, and train only a linear readout. Each control targets one ingredient reservoir theory requires: $\beta$ sets how strongly each new input is written in without overwriting what the state already carries, $\lambda$ generates nonlinear mixing of past inputs, and $\gamma$ provides the controlled forgetting on which fading memory depends.

\subsection{Configuration}
Unless stated otherwise, the reservoir has four qubits and two layers with ring coupling, uniform input injection, and amplitude damping; the readout is ridge regression on single-qubit $Z$ and ring-pair $ZZ$ expectations alone. We sweep the three controls $\beta$, $\lambda$, $\gamma$ on a fixed grid and select every hyperparameter, including the ridge penalty, on validation data only; the holdout never defines the operating region. Crossing the four benchmark tasks below with twenty frozen-reservoir seeds yields a balanced panel of $80$ task-seed replicates per grid coordinate. All grid coordinates and task–seed replicates are fixed before performance is inspected.

\subsection{Tasks, Metric, and Splits}
We use four scalar one-step-ahead prediction tasks spanning distinct demands: the Mackey--Glass and Lorenz systems for smooth chaos \cite{mackey1977,lorenz1963}, NARMA for nonlinear memory \cite{narendra1990}, and annual sunspot counts for irregular real-world data. Inputs are scaled to $[-1,1]$ and targets standardized; each task is split chronologically into washout, train, validation, and holdout segments. We score with normalized mean-squared error (NMSE), the mean squared error divided by the target variance, so that predicting the mean is the no-skill baseline and lower is better. NMSE is computed separately on validation and holdout.

\subsection{Operating Band}
The four tasks produce errors on very different scales, so raw NMSE values cannot be pooled across them; following standard practice for comparisons across data sets~\cite{demsar2006}, we compare settings by rank instead. Within each task $j$ and seed $s$, a setting $\theta\in\Theta_{\rm grid}$ receives the validation percentile rank $r_{j,s}(\theta)$, and lower is better. To guard against flukes---one easy task or one lucky seed---we keep every setting that frequently reaches a high rank rather than crowning a single winner. The operating band contains the settings that land in the top fraction $p$ for at least a fraction $q$ of task-seed pairs,
\begin{equation}
B_{p,q}=\{\theta:\Pr_{j,s}[r_{j,s}(\theta)\le p]\ge q\}.
\end{equation}
We call $B_{p,q}$ the band, and the dynamical behavior it identifies the operating regime. This frequency criterion mirrors stability selection~\cite{meinshausen2010} and treats the band as a search prior over settings that reliably work. We report the strict band $B_{20,0.7}$, the settings that reach top-quintile rank on at least $70\%$ of the task-seed replicates. We fixed these thresholds before the analysis, and the band is not sensitive to them: relaxing either one enlarges the same region rather than revealing new ones, from \PhaseBTwentyQSeventySize{} grid coordinates at $B_{20,0.7}$ to \PhaseBTwentyQSixtySize{} at $B_{20,0.6}$ and \PhaseBThirtyQSeventySize{} at $B_{30,0.7}$. In the transfer tests the band is built from the remaining replicates and scored on the withheld one; in the holdout audit it is frozen on validation, then evaluated once on holdout. Because ranks are formed within each task-seed replicate before aggregation, no large-error task or difficult seed can dominate the band.

\section{Results}
\begin{figure}[t!]
\centering
\includegraphics[width=\columnwidth]{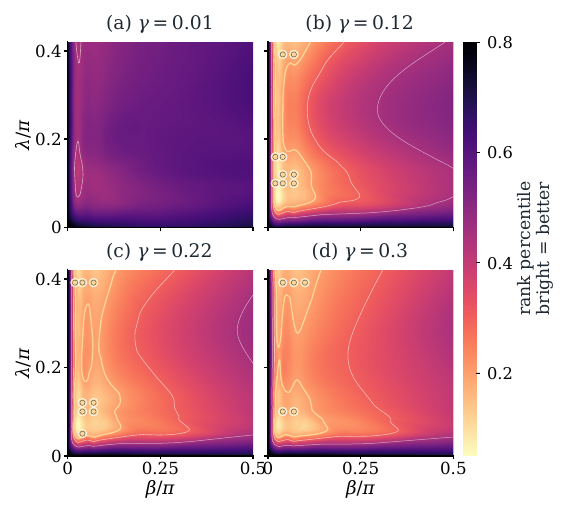}
\caption{The band requires nonzero damping. Each panel maps validation rank over $(\beta,\lambda)$ at the damping rate $\gamma$ named in its title (brighter $=$ better); gold circles are the $B_{20,0.7}$ band points at that $\gamma$. The band is essentially empty at $\gamma=0.01$ and fills in from $\gamma=0.12$ to $0.30$.}
\label{fig:gamma-slices}
\end{figure}

Settings with low validation rank form a stable operating band rather than scattered optima, and the band barely moves when any single task is dropped (Fig.~\ref{fig:phase}). The damping coordinate is selected by validation: the near-zero slice $\gamma=0.01$ contains almost no band points, while the strict $B_{20,0.7}$ core fills in from $\gamma=0.12$ to $0.30$ with medoid $\gamma=\PhaseRepresentativeGamma$ (Fig.~\ref{fig:gamma-slices}). Fig.~\ref{fig:phase} plots $\gamma=0.12$, the first populated slice; the medoid slice appears in Fig.~\ref{fig:gamma-slices}(c).

\begin{figure}[t!]
\centering
\includegraphics[width=\columnwidth]{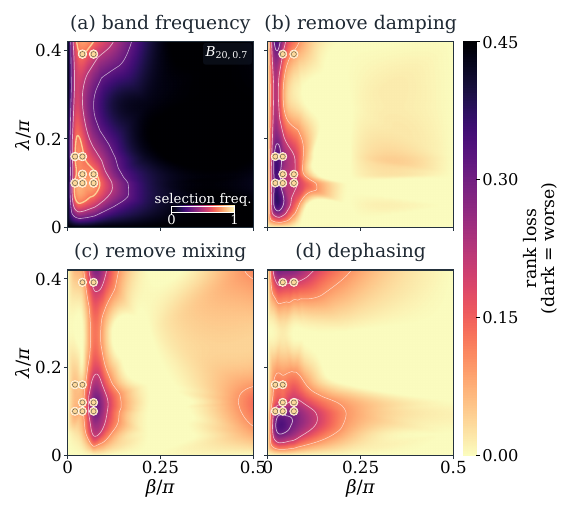}
\caption{Mechanism ablations destroy the operating band. (a)~Selection frequency in the original reservoir: for each coordinate $(\beta,\lambda)$, the color shows how often it reaches top-quintile validation rank across the 80 task--seed replicates; gold circles mark the $B_{20,0.7}$ band. (b)--(d)~Validation-rank loss for three representative ablations named in the panels, with darker colors indicating larger deterioration. The strongest losses occur at the original band coordinates.}
\label{fig:evidence}
\end{figure}

To test generalization beyond a joint fit, we rebuild the band with part of the data withheld and score it on exactly that part. When one task is withheld, the band built from the rest scores mean held-out rank \PhaseLotoMeanRank{}, 95\% bootstrap CI $[\PhaseLotoCILow,\PhaseLotoCIHigh]$; when one seed is withheld, \PhaseLosoMeanRank{}, 95\% bootstrap CI $[\PhaseLosoCILow,\PhaseLosoCIHigh]$. With $0.5$ the chance level, both values sit deep in the top quartile: the band transfers across task identity and reservoir initialization alike. It also holds on unseen data: frozen on validation and evaluated once on the chronological holdout, it remains in the top quartile on every task (Table~\ref{tab:holdout}), including NARMA, whose large raw NMSE reflects task difficulty rather than band failure. The regime, though not its coordinates, also survives an architectural change. A four-qubit, three-layer variant forms a band with similar mean held-out performance ($0.154$ vs.\ $0.153$), yet the two bands share almost no grid coordinates (Jaccard overlap $0.04$). Coordinate-level transfer thus fails, as expected once an added layer changes the per-step dynamics; what carries over is the regime, and the band need only be relocated for each architecture, which the diagnostic below makes cheap.
To test whether the band reflects the reservoir mechanisms rather than a coincidental grid location, we repeat the selection after ablating one ingredient at a time: no damping, no recurrent mixer, dephasing or depolarizing noise in place of amplitude damping, or $Z$-only\slash $ZZ$-only readouts. Figure~\ref{fig:evidence} shows three representative cases (no damping, no mixer, dephasing); the rest run under the same criterion but are not plotted. In the base model, the plotted $\gamma=0.12$ slice contains nine band coordinates with mean rank $0.223$, well inside the top quartile. Under every ablation, however, the re-selected band is empty: no coordinate is chosen often enough to satisfy the same $B_{20,0.7}$ stability criterion.

The original band coordinates also lose their advantage once the mechanisms are removed. Their mean rank deteriorates to $0.473$ without damping, $0.445$ with dephasing, $0.509$ with depolarizing noise, $0.367$ without the recurrent mixer, $0.486$ with $Z$-only readout features, and $0.468$ with $ZZ$-only features. Even the mildest case falls outside the top quartile, and most sit near chance. This rank loss concentrates exactly where the base band sat (Fig.~\ref{fig:evidence}): the ablations break precisely the region that was useful. Together these results indicate the regime emerges from the joint action of write-without-overwrite input, nonlinear mixing, controlled forgetting, and a readout accessing both local and pairwise observables.

\begin{figure*}[t!]
\centering
\includegraphics[width=\textwidth]{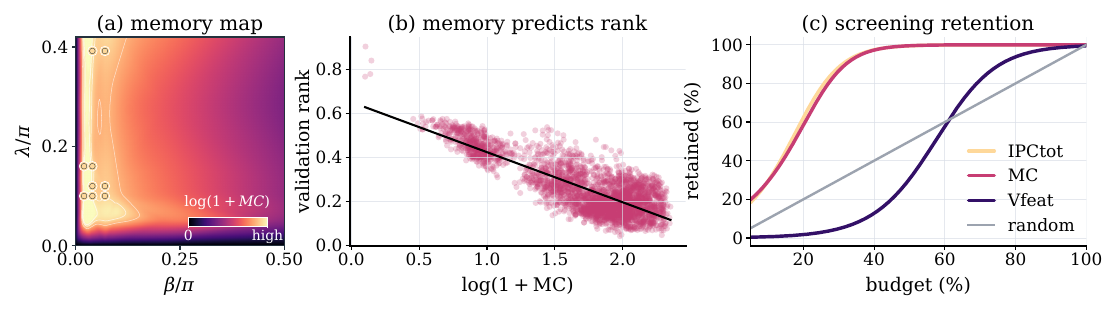}
\caption{Memory alone locates the band before any task is run. (a)~Task-free memory-capacity map, $\log(1+\mathrm{MC})$, on the $\gamma=0.12$ slice of Fig.~\ref{fig:phase}; gold points lie in the memory-rich region. (b)~Higher memory predicts lower validation rank (Spearman $\rho_s=-0.88$). (c)~Screening by each diagnostic's top-ranked fraction: $\mathrm{MC}$ and information-processing capacity ($\mathrm{IPC}$) retain top-decile validation-ranked pairs far faster than random selection, whereas raw feature diversity ($V_{\mathrm{feat}}$) barely beats chance.}
\label{fig:diagnostics}
\end{figure*}

Finally, the regime can be located before any task is run. We compute memory capacity from an independent random drive by fitting linear probes to reconstruct delayed inputs, clipping negative held-out $R^2$ values to zero, and summing across delays; information-processing capacity uses polynomial functions of the input history \cite{dambre2012}. Each probe uses the first $70\%$ of the post-washout diagnostic sequence for fitting and the remaining $30\%$ for scoring, and memory capacity sums delays 1--10. No benchmark-task labels enter these diagnostics. To use the score predictively, we rank seed--coordinate pairs by the diagnostic and retain the highest-scoring fraction. Figure~\ref{fig:diagnostics}(c) reports the fraction of top-decile validation-ranked pairs retained at each budget. The task-free memory map reproduces the band (Fig.~\ref{fig:diagnostics}a), and memory strongly predicts validation rank across settings and seeds ($\rho_s=\PhaseMCSpearman{}$); information-processing capacity behaves similarly, whereas raw feature diversity does not (Fig.~\ref{fig:diagnostics}b,c). Thus memory provides an operational screening tool before task-specific validation.


\begin{table}[t]
\centering
\scriptsize
\caption{Holdout performance after validation-only band selection. Rank is normalized so that $0.5$ is chance and lower is better.}
\label{tab:holdout}
\setlength{\tabcolsep}{3.2pt}
\begin{tabular}{@{}lccc@{}}
\toprule
Task & Band NMSE & Medoid NMSE & Band rank\\
\midrule
\PhaseHoldoutRows
\bottomrule
\end{tabular}
\end{table}
\section{Discussion}
For this canonical reservoir family, useful dynamics form a transferable regime without task-specific tuning. The regime persists when any task or seed is withheld. Memory, unlike raw feature diversity, identifies it. Reservoir theory explains why: drive, coupling, and dissipation must balance memory and nonlinearity \cite{tanaka2019,dambre2012}; each ablation collapses it.

We study simulated four-qubit realizations of a canonical dissipative QRC architecture that serves as a representative testbed; scaling with system size, finite-shot effects, calibration drift, device noise, and readout cost remain open \cite{mujal2023}. Because the readout features are low-weight Pauli expectations, shot noise enters at the standard $1/\sqrt{S}$ rate; testing the band on larger hardware is the natural next step. At this size the reservoir is also classically simulable, so the contribution is methodological, not a claim of quantum advantage. Within these limits, the practical consequence is concrete. The default workflow in QRC, a fresh hyperparameter search for every task, treats each problem as if the reservoir had to be rediscovered from scratch; our results say it does not: one memory measurement, taken once and without any task data, identifies the settings that then serve every task we tested. For the architectures and tasks tested, reservoir usefulness depends strongly on where it is operated, and promising settings can be screened from memory alone.

\section*{Code and Data Availability}
The \QRCprojectlink{} contains everything needed for full reproduction: simulator code, fixed grids, seeds, checked-in results, and figure scripts.

\section*{Acknowledgment}
This work was supported by the LMU Sustainability Fund (EfOiE), the German Federal Ministry of Research, \mbox{Technology and Space} (BMFTR) through QuCUN, QuaRDS, and CAQAO, the Munich Quantum Valley consortia K5 and K7, and the Bavarian Ministry of Economic Affairs project 6GQT.

\end{document}